\documentclass[english,twoside,a4paper,10pt]{article}

\usepackage[latin1]{inputenc}
\usepackage[T1]{fontenc}
\usepackage[english]{babel}
\usepackage{amsmath}
\usepackage{amsfonts}
\usepackage{graphicx}
\usepackage{subfigure}
\usepackage{a4wide}
\usepackage{amssymb}
\usepackage{fancyhdr}
\usepackage{mathrsfs}
\usepackage[toc,page]{appendix}

\def\ba{\begin{eqnarray}}
\def\ea{\end{eqnarray}}

\def\be{\begin{equation}}
\def\ee{\end{equation}}

\begin{document}

\title{{\bf Higher derivative and mimetic  models on non flat  FLRW space-times}}

\author{ 
Alessandro Casalino$^1$\footnote{E-mail address: alessandro.casalino@unitn.it},\,\,\,
  Lorenzo Sebastiani$^2$\footnote{E-mail address: lorenzo.sebastiani@unitn.it},\,\,\,
Luciano Vanzo$^1$\footnote{E-mail address: luciano.vanzo@unitn.it},\,\,\,
  Sergio Zerbini$^1$\footnote{E-mail address: sergio.zerbini@unitn.it}\\
\\
\begin{small}
$^1$Dipartimento di Fisica, Universit\`a di Trento, Via Sommarive 14, 38123 Povo (TN), Italy
\end{small}\\
\begin{small}
TIFPA - INFN,  Via Sommarive 14, 38123 Povo (TN), Italy
\end{small}\\
\begin{small}
$^2$ Istituto Nazionale di Fisica Nucleare, Sezione di Pisa, Italy
\end{small}\\
\begin{small}
Dipartimento di Fisica, Universit\'a di Pisa, Largo B. Pontecorvo 3, 56127 Pisa, Italy.
\end{small}
}

\maketitle
\abstract{An effective Lagrangian approach, partly inspired by Quantum Loop Cosmology (QLC), is presented and formulated in a  non flat  FLRW space-times, making use of modified gravitational models. The  models considered are non generic, and their choice is dictated by the necessity to have at least second order differential equations of motion in a non flat FLRW space-time. This is accomplished by a class of Lagrangian which are not analytic in the curvature invariants, or making use of a mimetic gravitational scalar field. In this paper we want to show that, for some effective models, although the associated generalized Friedmann equation may admit non singular metrics as solutions, the de Sitter space-time is present only for a restricted class, which includes General Relativity and Lovelock gravity. The other models admit pseudo de Sitter solutions, namely FLRW metrics, such that for vanishing  spatial curvature looks like flat de Sitter patch, but for non vanishing spatial curvature are regular  bounce metrics or, when the spatial curvature is negative, Big-Bang  singular metrics.  
}

\section{Introduction}

The theory of General Relativity (GR), with a suitable cosmological positive constant and the addition of the dark matter, is in accordance with the experiments in a vast part of the history of the Universe, including the acceleration (or Dark Energy dominated) era.  With an additional scalar degree of freedom, this model can also describe the primordial inflationary period, which is a possible solution for the horizon and flatness problem. This is essentially the so called $\Lambda$CDM model, or the standard cosmological model, and it has been recently tested with high accuracy \cite{Plank1,Planck2, Plank2}.  With regard to this, very recently, the evidence also for a possible closed universe has been reported \cite{DiVa}. 

It is also well known that the field equations of GR are a system of partial (quasi-linear) differential equations which are at most at the second order in the derivatives. This latter property has profound mathematical and physical implications. For instance, this fact is a necessary condition in order to build a consistent Hamiltonian formulation.

In a classical theory of gravity, a celebrated theorem due to Lovelock \cite{Lov} states that in order to have second order partial differential equations as equation of motions for a scalar Lagrangian depending only on curvature invariants, we have to deal with higher dimensional space-times. In $D=4$, the only Lagrangian admitting second order differential equation is the Einstein-Hilbert Lagrangian, up to the addition a cosmological constant. Both GR and Lovelock gravitational theories admit solutions corresponding to singular space-times, namely metrics whose scalar curvature invariants have singularities, or equivalently there exist geodesic incomplete metrics. 

It is believed that quantum corrections to GR might solve the aforementioned singularities issues. In fact, in Quantum Loop Cosmology (QLC), an effective modified Friedman equation has been obtained, whose solution in a flat FLRW space-times admits a \textit{bounce}, a solution without the GR Big Bang singularity \cite{Bo,Bo2}. See also Refs. \cite{Bo19,da19}.

On the light of reference \cite{DiVa}, our aim in this paper is to extend this effective approach, inspired by QLC, to non flat FLRW space-times, making use of modified gravitational models. We will show that for some effective models, the associated generalized Friedmann equation, in general, may admit non singular FLRW  metrics as solutions, but there exist pseudo de Sitter space-time  solutions, namely  metrics which coincide with the flat FLRW de Sitter patch, the one with vanishing spatial curvature.  For non vanishing spatial curvature, and in the presence of a positive cosmological constant, the solutions are regular bounces for positive spatial curvature or singular Big-Bang metrics for negative spatial curvature.

The fact that the generalized Friedmann equations admit pseudo  dS solutions might partially be understood observing that some of modified gravitational models are described by Lagrangians which are non-analytic in the curvature invariants, becoming singular on  dS space-times. 

The paper is organized as follow. In Section 2, a brief review concerning dynamical spherical symmetric space-times is presented. In Section 3 and 4 we discuss the issue associated with the dS solution respectively for some mimetic models and non-polynomial gravity models. In Section 5 we present an example of the aforementioned dS problem in the context of Horndeski gravity. In Section 6 we review Lovelock gravity and we show that these models are not affected by the pseudo dS issue. In Section 7 we draw some conclusions.

If not otherwise stated, in this paper we will consider the convention $8\pi G_N=1$.

\section{Spherically symmetric space}

The $D$-dimensional Friedmann-Lemaitre-Robertson-Walker (FLRW) space-times are an example of Spherically Symmetric Space-times (SSS). For the sake of completeness, we shall briefly review the formalism we are going to deal with in the following sections. We follow Refs. \cite{noi,noi2}.  

The generic SSS  metric reads
\begin{align}
ds^2=g_{\mu \nu }dx^\mu dx^\nu =\gamma_{ab}(x)d x^a dx^b +r^2(x)dS^2_{D-2}\,,
\label{eq:ansatz}
\end{align} 
where $a,b = 0, 1$ and $r$ is a scalar quantity, while $S^2_{D-2}$ is the $D-2$ dimensional sphere.

Other relevant scalar quantities on SSS are
\be
\chi=\gamma^{ab}\partial r_a \partial r_b\,,
\label{chi}
\ee
and
\be
\Phi=\nabla^2_\gamma  r\,,
\label{lap}
\ee
where $\nabla^2_\gamma$ is the two dimensional Laplacian on the two dimensional normal space-time whose metric is $\gamma_{ab}$.

For example, in a non flat FLRW space-time, the metric being
\begin{equation}
ds^{2}=-dt^{2}+a(t)^2\left(
\frac{d\rho^2}{1-k\rho^2}+\rho^2dS^2_{D-2}
\right)\,,\quad k=0,\pm K_0\,,
\label{metric}
\end{equation}
where $a\equiv a(t)$ is a function of the time only,
we obtain  $x=(t,\rho)$,  $r=a(t)\rho$, $\gamma_{ab}=\mbox{diag}\left(-1, \frac{a^2}{1-k\rho^2}\right)$, and we get
\be
\chi=\gamma^{ab}\partial r_a \partial r_b=1-r^2 \, J^2\,,\quad J^2=H^2+\frac{k}{a^2}\,,
\label{chi1}
\ee
where $H=\frac{\dot{a}}{a}$ is the Hubble parameter. The other invariant (\ref{lap})
reads
\be
\Phi=\nabla^2_\gamma  r= -(J^2+Q^2) r\,, \quad Q^2=H^2+\dot{H}\,,
\label{lap2}
\ee
the dot being the derivative with respect to the time coordinate $t$.
As a result, $J^2$ and $Q^2$ are confirmed to be scalar quantities in a generic SSS. It should be noted that the Ricci scalar is given by
\be
R=6\left[J^2+Q^2\right]\,.
\ee
Now we see  an example, which it will be useful later, namely the
 de Sitter (dS) space-time. Besides the static patch
\be
ds^2=-(1-H_0^2r^2)dt^2+\frac{dr^2}{1-H_0^2r^2}+r^2dS^2\,,
\ee
where $H_0$ is a constant,
dS space admits three FLRW space-times patches.

The first one is flat FLRW patch with $k=0$, 
\begin{equation}
ds^{2}=-dt^{2}+e^{2H_0t}\left(d\rho^2+\rho^2dS^2_{2} \right)\,,
\label{ds0}
\end{equation}
where $a(t)=e^{H_0t}$, and $H=H_0$. From equation \eqref{chi1}, we obtain $J^2=H_0^2=Q^2$.

Then we have the $k>0$ patch. With a set of new coordinates $(T,R)$ it can be written as
\be
ds^2=-dT^2+\cosh^2H_0T \left(\frac{dR^2}{1-H_0^2R^2}+R^2dS_2 \right)\,.
\label{1}
\ee
Here, $a(T)=\cosh H_0 T$, $k=H_0^2$ and $H=H_0\frac{\sinh H_0 T}{\cosh H_0 T}$. We still have $J^2=H^2_0=Q^2$, as in the flat case. 

Finally, we have the $k<0$ dS patch. With another set of coordinates $(\tau,\sigma)$ we obtain,
\be
ds^2=-d\tau^2+\sinh^2H_0\tau \left(\frac{d\sigma^2}{1+H_0^2\sigma^2}+\sigma^2dS_2 \right)\,,
\label{-1}
\ee
where $a(\tau)=\sinh H_0\tau $, $k=-H_0^2$ and $H=H_0\frac{\cosh H_0 \tau}{\sinh H_0 \tau}$. Again, we have $J^2=H^2_0=Q^2$. In all dS patches, the Ricci scalar is constant and reads $R=12H^2_0$

We conclude this Section by recalling that there exists a dynamical trapping horizon (see for example Refs. \cite{noi,noi2} and references quoted therein) when $\chi_H = 0$ (for instance, in FLRW when $r_H^2J^2=1$). In this case, there is a related dynamical surface gravity, dubbed Hayward surface gravity, given by
\be
k_H=\frac{1}{2}\Phi_H=\frac{1}{2}\left(\nabla^2_\gamma  r\right)_H= -\frac{\left[J^2+Q^2\right]}{2 }r_H\,.\label{h}
\ee
It is easy to show that the dynamical horizon for dS space-times in all the versions reads $r_H=\frac{1}{H_0}$, and a surface gravity  $|k_H|=H_0$.
The possible relation of $k_H$ with a dynamical Hawking effect and Hawking temperature is discussed in Refs. \cite{noi,noi2,noi1,zocc}.

\section{Mimetic models}

If we want to work in $D=4$, preserving the GR property of having second order differential equations in the FLRW space-time, we might consider the so called Horndeski models \cite{horn}, Horndeski mimetic models \cite{muk,muk1,vik,Derue,barvi,beke,mata,golo,Lim,Capo,Odi1,Odi2,Odi3,myrza1,myrza2,myrza3,mim_sv1,mim_sv2},
DOHST models \cite{Deffa,DeFe}, or Non Polynomial Gravity models \cite{aimeric, aimeric2}. In this Section we present the issue associated with the dS solution for some mimetic models.

\subsection{A covariant renormalizable model}

We firstly consider the following mimetic scalar tensor gravity model \cite{NO,Cogno11}, 
\be
I =  \int_{\mathcal M} d^4 x \sqrt{-g} \left[\frac{R-2\Lambda}{2} + \lambda\left(X-\frac{1}{2}\right)  -V(\phi) \right]+I_H+I_m\,,
\label{d}
\ee
where $X=-\frac{1}{2}g^{\mu \nu}\partial_\mu \phi \partial_\nu \phi$,  $\lambda$ is a Lagrange multiplier, $\phi$ is the mimetic scalar field whose potential is $V\equiv V(\phi)$, $\Lambda$ the cosmological constant and $I_m$ is the matter-radiation action of a perfect fluid. The higher order contribution is given by
\be
I_H=\int_{\mathcal M} d^4 x \sqrt{-g}\,\,\alpha( G_{\mu \nu} \nabla^\mu \phi  \nabla^\nu \phi)^n\,,
\label{h2}
\ee
where $\alpha$ and $n$ are constants. The above Lagrangian is a particular case of the general mimetic  Horndeski Lagrangian \cite{horn,Deffa,DeFe}. Some examples of Horndeski mimetic gravity models have  been recently considered in Ref. \cite{mata}. When the constant $\alpha$ is vanishing, and in the absence of matter, the above model reduces to the original mimetic gravity proposed by Chamseddine and Mukhanov \cite{muk1}. In the following, we put $V(\phi)=0$ and we omit the matter contributions.

In order to compute the field equations, we use the line element
\begin{equation}
ds^{2}=-N^2(t)dt^{2}+a(t)^2\left(
\frac{dr^2}{1-kr^2}+r^2d S^2 \right)\,\,,
\label{metricn}
\end{equation} 
where $N\equiv N(t)$  is an arbitrary dynamical variable which takes the value $N=1$ after its variations. The action is therefore a functional of $a(t)$, $N(t)$ and $\lambda$. Assuming an homogeneous and isotropic scalar field $\phi = \phi(t)$, i.e. a dependency on $t$ only, the variation with respect to $\lambda$ gives the mimetic constraint
\be
\dot{\phi}^2=1\,.
\label{l}
\ee
Thus, in the following we can consider $\phi=t$. From the variation with respect to $N$ (and considering $N=1$ after the variation), we obtain the generalized Friedmann equation, which reads
\be
6 J^2 -2\Lambda -\alpha(-3)^nJ^{2n} +2\alpha(-3)^nJ^{2n-2} H^2 =\lambda  \,,
\label{f1}
\ee
where we have introduced the scalar quantity $J^2$ defined in (\ref{chi1}).
Then, the equation of motion associated with $\phi$ gives
\be
\lambda=-2n\alpha (-3)^nJ^{2n}\,.
\ee
As a result, the Lagrangian multiplier can be eliminated and the final equation reads
\be
6 J^2 +(2n-1)\alpha(-3)^nJ^{2n} +2\alpha(-3)^nJ^{2n-2} H^2 =2\Lambda  \,.
\label{f3}
\ee
As we noted before, for $n=1$ we are considering a mimetic Hordenski gravity model. Let us investigate this case, namely  
\be
6 J^2 -3\alpha J^{2} -6\alpha  H^2 =2\Lambda  \,.
\label{f4}
\ee
It should be noted that in this equation the quantities $J^2$ and $H^2$ appear. Therefore, for $k=0$, $J^2=H^2$ and the dS-like solution with  $H=H_0$ exists. However, the other two dS patches are not solutions of the above equation, due to the fact that $J^2$ should be again $J^2=H_0^2$, namely a constant, but the term $H^2$ explicitly depends on the time. 

Other solutions can be found for $k\neq 0$. As a differential equation for $a(t)$, we find
\be
(6 -9\alpha) \dot{a}^2 -2\Lambda a^2 =-k(6-3\alpha)  \,.
\label{f44}
\ee
Let us assume positive $\Lambda$ and $0\leq\alpha <\frac{2}{3}$. Thus, for $k=0$, we get
\be
a(t)=e^{H_0 t}\, , \quad H^2_0=\frac{2\Lambda}{6-9\alpha}\,.
\label{0}
\ee
However, for non vanishing spatial curvature, there exist the two following solutions
\be
a(t)=\cosh H_0 t\,, \quad  k=\mu^2>0\,,\nonumber
\ee
\be
a(t)=\sinh H_0 t\,, \quad  k=-\mu^2 <0\,,\label{sol00}
\ee
where
\be
\mu^2=H_0^2\frac{6-9\alpha}{6-3\alpha}=\frac{2\Lambda}{6-3\alpha}\,.\label{eq:mu2}
\ee
The first one is a regular bounce solution, with $J^2$ finite but non-constant. The second one is a Big-Bang singular solution at $t=0$.

From Eq. \eqref{eq:mu2}, we see that the only way to obtain $\mu^2 = H_0^2$, as required by the de Sitter form of the metric \eqref{1} and \eqref{-1}, is to consider $\alpha=0$, i.e. GR. This is not mysterious at all since de Sitter space is a maximally symmetric vacuum solution of Einstein equations. A static cosmological horizon with $r_H=1/J=1/H_0$ is associated to the dS space-time (see the last paragraph of the preceding Section).
Here,
when $\alpha\neq 0$ and $k\neq 0$, we get a dynamical cosmological horizon. In the specific, for $k>0$  we obtain $1/H_0^2\leq r_H^2\leq (6-3\alpha)/(H_0^2(6-9\alpha))$ with $r_H^2(t\rightarrow\pm\infty)=1/H_0^2$ and $r_H^2(t=0)=(6-3\alpha)/(H_0^2(6-9\alpha))$.
On the other side, when $k<0$, $0\leq r_H^2\leq 1/H_0^2$ with $r_H^2(t=0)=0$ 
and $r_H^2(t\rightarrow\infty)=H_0^2$. 

As a consequence, the dS solution does not exists in this model, but there exists three distinct cosmological FLRW models associated with the value of spatial curvature.

\subsection{An extended mimetic gravitational model} 

In this Section we revisit the extended mimetic model introduced in Ref. \cite{Muk3} by making use of a Lagrangian minisuperspace approach. Here, however, we shall deal with non flat FLRW space-times, as considered very recently in Ref. \cite{Muk4} within a different approach.

We start with the action
\be
I =  \int_{\mathcal M} d^4 x \sqrt{-g} \left[\frac{R}{2} + \lambda\left(X-\frac{1}{2}\right)+f[\chi(\phi)] \right]+I_m\,.
\label{d_m}
\ee
The higher order differential term in $\phi$, $f[\chi(\phi)]$,  depends on $\chi(\phi)=-\nabla^\mu \nabla_\mu \phi \, /3$.

With the metric \eqref{metricn}, the action is again a functional of $a(t)$, $N(t)$ and $\lambda$. However, we may simplify the derivation of the equations of motion considering  $N(t)=1$, and again assuming $\phi = \phi(t)$, i.e. a dependency on $t$ only. The variation with respect to $\lambda$ gives again the mimetic constraint \eqref{l}. Therefore in the following we can consider $\phi=t$. Thus 
\begin{equation}
\chi(\phi) = -\frac13 \nabla^\mu \nabla_\mu \phi =- \frac13 \partial^\mu \partial_\mu \phi + g^{\mu \nu} \frac13 \Gamma_{\mu \nu}^\rho \partial_\rho \phi = H\,.
\end{equation}
For our purpose we would like to consider the non-vacuum case, namely now we will take into account the contribution of standard matter. In this case, it is convenient to start with
the variation of the action with respect to the scale factor $a$, which leaves to the generalized second Friedmann equation, 
\be
 J^2(t)+2Q^2(t)+\frac{f(H)}{2}-\frac{H}{2}\frac{d f(H)}{dH}-\frac16 \frac{d}{dt}\frac{d f(H)}{dH}  =-p  \,,
\label{z1}
\ee 
$p$ being the matter pressure.
Using the matter conservation law
\be
\dot{\rho}=-3H(\rho+p)\,,\label{matter_cons}
\ee
where $\rho$ is the matter energy density, 
we can easily obtain the generalized first Friedman equation as
\be
 6J^2(t)+f(H)-H\frac{d f(H)}{dH}  =2\rho  \,.
\label{z2}
\ee 
Furthermore, let us consider the  choice \cite{Muk3,Lang,Haro}
\begin{equation}
 f(H)=6H^2+\frac{12}{\alpha^2}\left[
 1-\sqrt{1-\alpha^2 H^2}-\alpha H\arcsin\left(\alpha H\right)
 \right]\,,
\end{equation}
where $\alpha$ is a dimensional positive parameter. Note that, since $f(H)$ goes to zero when $\alpha\rightarrow 0$, in this limit one recovers GR. Thus $f(H)$ may represent a ``correction'' to pure mimetic gravity. Obviously, we must require $H^2<1/\alpha^2$.

From equation (\ref{z2}) we find
\begin{equation}
\frac{6}{\alpha^2}\left[
1-\sqrt{1-\alpha^2 H^2}\right]=\rho -\frac{3 k}{a^2}\,,\label{EOM1bis}
\end{equation}
which is equivalent to 
\begin{equation}
3H^2=\left(\rho-\frac{3k}{a^2}\right)\left(1-\frac{\rho-\frac{3k}{a^2}}{\rho_c}\right)\,,\qquad \text{where}\qquad  \rho_c=\frac{12}{ \alpha^2}\,.
\label{zf}
\end{equation}
In QLC (see \cite{Lang} for a review), $\rho_c$ is proportional to the Planck energy density $\rho_{Pl}$ and, analogously to the flat QLC case, in general $0<\rho-3k/a^2 \ll \rho_c$. Therefore, when $k\neq 0$, $\rho_c$ can not be considered as a critical density. For an alternative Lagrangian derivation within a mimetic approach and when $k$ is not vanishing, see Ref. \cite{Lang} and references therein.

For $k=0$ we obtain
\begin{equation}
3H^2=\rho\left(1-\frac{\rho}{\rho_c}\right)\,.\label{aboveeq}
\end{equation}
This is the  QLC modified Friedmann equation in flat FLRW space-time. For an equation of state $p=\omega \rho$,
$\omega\neq -1$,
it  admits a well known bounce solution 
\begin{equation}
a(t)=\left[\frac{\rho_0}{\rho_c}+\frac{3\rho_0(1+\omega)^2}{4}\,t^2  \right]^{1/[3(1+\omega)]}\,.
\label{bflat}
\end{equation}
When the critical density  $\rho_c$ is going to infinity, we recover the GR Big-Bang solution. For other cosmological bounce solutions see Ref. \cite{Haro,Biswas1,Biswas2} and references therein.

Furthermore, in the case $\omega=-1$, namely $\rho=\rho_0$, the above equation (\ref{aboveeq}) admits a $k=0$ dS solution. 

On the other side, when $k \neq 0$, we should use of Eq. (\ref{zf}) which can be rewritten as
\begin{equation}
3 J^2=\rho-\frac{\left(\rho-\frac{3k}{a^2}\right)^2}{\rho_c}  \,.
\end{equation}
Then using $J^2=H^2+ka^{-2}$, and for $k \neq 0$, this equation does not admit in general a pure (vacuum) dS solution. Instead, we have the one-parameter family of particular solutions for $\omega=-1$ and $\rho=\rho_0$,
\be
a^2(t)=\frac{3k(\rho_c-2\rho_0)}{2\rho_0(\rho_c-\rho_0)}+a_0 \cosh(H_0 \, t)\, ,\quad H_0^2=\frac{4 \rho_0 (\rho_c-\rho_0)}{3 \rho_c}\,,
\ee
with
\be
a_0^2=\frac{9}{4}\frac{k^2 \rho_c^2}{\rho_0^2(\rho_0-\rho_c)^2}\,.
\ee
As a result, if $\rho_0\ll \rho_c$ for $k>0$, we find a bounce solution
and we see that the qualitative behaviour is not far from de Sitter space. However, we note that in principle $\rho_0$ may be larger than $\rho_c$ while it still satisfies the requirement $\rho_0-3k/a^2\ll\rho_c$. In this case, the bounce solution is not present.

For $k<0$, the Big Bang singularity appears. Since $\rho_0 \ll \rho_c$ (such that $\rho_0-3k/a^2\ll \rho_c$), we get
\begin{equation}
H(t)\simeq \frac{H_0}{2}\frac{\sinh H_0 t}{\cosh H_0 t - 1}\,\quad \text{for} \quad k<0\,, 
\end{equation}
and the solution diverges when $t\rightarrow 0$. Thus, in this case, $H$ can become arbitrarily large, and the condition $H^2<1/\alpha^2 = \rho_c / 12$ is satisfied in the QLC case for times where $H(t) \lesssim \sqrt{\rho_{Pl}}$.

\subsection{Non vanishing spatial curvature models}

When $k \neq 0$, and in presence of matter, in general the Big -Bang singularity may be absent. First we give an example of exact solution for the extended mimetic model presented above. Consider the barotropic equation of state  $p=-\rho/3$, i.e. the equation of state parameter is $\omega= -1/3$. As a result,  $\rho(t)=\rho_0 a(t)^{-2}$. It is convenient to introduce the quantity $y(t)=a(t)^2$. In this case, equation (\ref{zf}) becomes
\begin{equation}
\frac{3}{4} \dot{y}^2=(\rho_0-3k)y -\frac{(\rho_0-3k)^2}{\rho_c} \,.
\label{zf1}
\end{equation}
The related solutions are
\be
y(t)=a^2(t)=\frac{\rho_0-3k}{\rho_c}+\left[C \pm \sqrt{\frac{\rho_0-3k}{3}  }\,t   \right]^2\,,
\ee
where $C$ is an arbitrary dimensionless constant of integration. Moreover we assume $\rho_0>3k$ in order to obtain a real solution. These are regular bounce solutions, with $a(0) \neq 0$. When $C=0$, the regular solutions become a unique symmetric bounce solution, namely
\be
a(t)^2= \frac{\rho_0-3k}{\rho_c}+\frac{\rho_0-3k}{3}  \,t^2  \,.
\ee
The related density is also regular. When $\rho_c$ goes to infinity, we recover the well known  GR solution, admitting  the Big Bang singularity.

For a generic $\omega$, it is not easy to find an exact solution. Alternatively, we may start separating the variable in equation (\ref{zf}) with $y=a^2$, namely
\be
\int \frac{dy}{\sqrt{Y(y)}}=t\,,
\ee
where we have used the matter conservation law and put
\be
Y(y)=\frac{4 \rho_0}{3}\left[y^{1/2-3\omega/2}-\frac{\rho_0}{\rho_c}\,y^{-(1+3\omega)}-\frac{3k\,y}{\rho_0} +
\frac{6k}{\rho_c}y^{-(1/2+3/2\omega)}-\frac{9k^2}{\rho_c\rho_0} \right]\,.
\label{zz}
\ee
If $k$ is not vanishing, the above integral is only exactly computable only for $\omega=-1/3$. 

However, if we make an expansion around the critical point defined by  $ Y(y_*)=0$, namely
\be
y^{1/2-3\omega/2}_*-\frac{\rho_0}{\rho_c}\,y^{-(1+3\omega)}_*-\frac{3k\,y}{\rho_0} +
\frac{6k}{\rho_c}y^{-(1/2+3/2\omega)}_*-\frac{9k^2}{\rho_c\rho_0} =0\,,
\label{z3}
\ee
we obtain the approximate solution, valid for small $t$
\be
y(t)\simeq y_*+\frac{Y'_*}{4}\,t^2\,,
\label{zz4}
\ee
where 
\be
Y'_*=\frac{4 \rho_0}{3}\left[  \frac{1-3\omega}{2}y^{-1/2-3\omega/2}+\frac{\rho_0(1+3\omega)}{\rho_c}\,y^{-(2+3\omega)}_*-\frac{3k}{\rho_0} -
\frac{3k(\frac{1+3\omega}{2} )}{\rho_c}y^{-(3/2+3/2\omega)}_* \right]\,.
\label{zzz}
\ee
Above, $y_*$ is solution of the transcendental equation 
(\ref{z3}).  It is easy to show that for $\omega= -1/3$, we obtain the approximate solution related to the exact solution found before.  Thus, from  equation (\ref{zz4}), we may conclude that we should deal with a regular symmetric bounce as soon as $y_*>0$ and $Y'_*>0$. 

As a check, for $k=0$, we obtain $y^{3/2+3/2\omega}=\frac{\rho_0}{\rho_c}$, thus $y_*>0$, and $Y'_*=\frac{4 \rho_0^2}{3\rho_c}(1+\omega)y_*^{-(2+3\omega)}>0$, in agreement with well known exact flat bounce solution. 

\section{Non-Polynomial gravity models}

The issue associated with non existence of the dS solution for mimetic models persists also in the case of Non-Polynomial (NP) gravity models, as we will show in this Section. In fact, we will find that the non-analyticity of these models might lead to the pseudo dS issue.

\subsection{The $F(R,G)$-model}

Consider a modified gravity model described by the following $F(R,G)$-gravity action,
\begin{eqnarray}
I = \int_\mathcal{M} d^4 x \sqrt{-g}\, \frac{F(R, G)}{2} + I_m \,,
\label{action2}
\end{eqnarray}
where, again, $I_m$ is the matter-radiation action of a perfect fluid, $F\equiv F(R,G)$ is a generic function of the Ricci scalar and the Gauss Bonnet four dimensional topological invariant defined as
\begin{equation}
G=R^{2}-4R_{\mu\nu}R^{\mu\nu}+R_{\mu\nu\xi\sigma}R^{\mu\nu\xi\sigma}\,.\label{GaussBonnet}
\end{equation}
If we consider a generic non flat FLRW space-time, and if we use again the line element \eqref{metricn}, then the Ricci scalar and the Gauss-Bonnet assume the form
\begin{eqnarray}
R&=& 6\left(
\frac{\ddot a}{a N^2}+\frac{\dot a ^2}{a^2 N^2}-\frac{\dot a \dot N}{a N^3}+\frac{k}{a^2}\right)\,,\label{R}\\
G&=& \frac{24 }{a^3 N^3}\left(
\frac{\dot a^2 \ddot a}{N}-\frac{\dot a^3\dot N}{N^2}-\dot a\dot N k+N\ddot a k 
\right)\label{G}\,.
\end{eqnarray}
Substituting these expressions inside the Lagrangian we obtain a higher derivative theory. However, in order to simplify the computation and the final expression, we can use two Lagrangian multipliers $\lambda$ and $\mu$ and thus rewrite the action as \cite{superSeba}
\begin{eqnarray}
I&=& \int dt N\,a^3 \,
\biggl\lbrace F(R,G)-
\nonumber\\&&
-\lambda\left[R-6\left(
\frac{\ddot a}{a N^2}+\frac{\dot a ^2}{a^2 N^2}-\frac{\dot a \dot N}{a N^3}+\frac{k}{a^2}\right)
\right]
\nonumber\\
&&\left.-\mu\left[G-\frac{24 }{a^3 N^3}\left(
\frac{\dot a^2 \ddot a}{N}-\frac{\dot a^3\dot N}{N^2}-\dot a\dot N k+N\ddot a k 
\right)
\right]\right\rbrace\, + I_m.
\end{eqnarray}
The variations with respect to $R$ and $G$ lead to
\begin{equation}
\lambda=F_R\,,\quad \mu=F_G\,. 
\end{equation}
Thus, after integration by parts, we obtain the Lagrangian in the gravitational sector
\begin{equation}
\mathcal{L}(N,a, R, G)=Na^3(F-RF_R-GF_G)-\frac{6\dot a^2 a F_R}{N}-\frac{8\dot a^3\dot F_G}{N^3}-\frac{6\dot a a^2\dot F_R}{N}+k\left(6Na F_R-\frac{24\dot a \dot F_G}{N}\right)\,, 
\end{equation}
which involves only first derivatives of the given variables.

The variations with respect to $N$ and $a$ yield
\begin{align}
&6 J^2F_R+\left(F-RF_R-G F_G\right)+6H\left(\dot F_R+4J^2\dot F_G\right)=2\rho\,,\\
&8H^2\ddot F_G+2\ddot F_R+4H\dot F_R+16H\dot F_G(\dot H+H^2)+F_R(4\dot H+6H^2)+(F-R F_R-G F_G)+\nonumber\\
&\phantom{xxxxxxxxxxxxxxxxxxxxxxxx}+\frac{k}{a^2}(2F_R+8\ddot F_G)=-2 p\,,
\end{align}
where we fixed $N=1$ after the variation, and we considered the contribution of matter. Obviously, for $F(R,G)=R$ we recover the GR Friedmann equations. 

Since this model is a higher derivative theory, the above equations of motion involve the presence of third and fourth order time derivatives of the dynamical variable $a(t)$. Furthermore, in general, the dS solution exists in all patches. For example with the choice
\be
F(R,G)=R+\frac{\alpha}{2}\,R^2+\frac{\beta}{2}\,G^2\,,
\ee
we find
\be
6J^2(1+\alpha R)-\frac{\alpha}{2}R^2-\frac{\beta}{2}G^2+6H\left[\alpha\dot R+4J^2\dot G\right]=2\rho\,.
\ee
In the dS case,  $R_0=12H_0^2$ and $G_0=24H_0^4$ are constant, and for all dS patches $J^2=H_0^2$. Thus, with $\rho_0$ constant or vanishing
\be
6 H_0^2=\frac{\beta}{2}G^2_0+ 2\rho_0\,,
\ee
which determines  $H_0$, namely the dS curvatures.  Note that this example is not affected by the dS issue.

We can obtain the same result starting from dS existence condition for $ F(R,G)$ modified gravity which reads \cite{Monica}
\begin{equation}
2F=RF_R+2GF_G\,.
\label{dsfg}    
\end{equation}

In the following, we would like to discuss another suitable choice for $F(R,G)$, such that the above equations of motion contain only first and second order time derivatives of $a(t)$.  We will follow Ref. \cite{Gao}.
Let us consider
\begin{equation}
F(R,G)=R+f\left[J^2(R,G)\right]\quad\text{with}\qquad J^2=\frac{R+\sqrt{R^2-6G}}{12}\,. 
\end{equation}
In fact, on a generic FLRW space-time, we find
\begin{equation}
J^2=H^2+\frac{k}{a^2}\,. 
\end{equation}
 Note that this choice necessarily contains a non analytic dependence on the variables $R$ and $G$.
The first modified Friedmann equation with this choice becomes
\begin{equation}
6J^2+f(J^2)- \frac{H^2}{J} \frac{\partial f}{\partial J}=2\rho\,. \label{EOM1}
\end{equation}
We may write the above equation in the form
\begin{equation}
  6J^2+f(J^2)- J \frac{\partial f}{\partial J}=2\rho-\frac{k}{a^2 J} \frac{\partial f}{\partial J} \,.
  \label{EOM2}
\end{equation}
This equation is similar to the one derived within the extended mimetic model. Thus, making again the choice 
\begin{equation}
 f(J)=6J^2+\frac{12}{\alpha^2}\left[
 1-\sqrt{1-\alpha^2 J^2}-\alpha J\arcsin\left(\alpha J\right)
 \right]\,,
\end{equation}
where $\alpha$ is again a dimensional positive parameter, we find 
\begin{equation}
\frac{6}{\alpha^2}\left[
  1-\sqrt{1-J^2\alpha^2}\right]=\rho -\frac{6 k}{a^2}\left(1-\frac{\arcsin (\alpha J)}{\alpha J} \right)\,.
\label{EOM1tris0}
\end{equation}
This leads to 
\be
3J^2=\rho-\rho_k(J)-\frac{[ \rho-\rho_k(J) ]^2}{\rho_c}\,, \quad \rho_c=\frac{12}{\alpha^2}\,,
\label{bu}
\ee
where
\be
\rho_k(J)=\frac{6 k}{a^2}\left[1-\frac{\arcsin (\alpha J)}{\alpha J} \right]\,.
\ee
When $k=0$, $\rho_k(J)=0$, and the above equation  coincides with the one obtained in the mimetic approach and, of course, they  have the same flat bounce FLRW bounce solution. Again, when $\rho=\rho_0$ constant, for $k=0$, we obtain a flat dS like FLRW solution. On the other side, for $k$ different from zero, dS solution does not exist. In this model, we may understand the non existence of the dS solution, since in the Lagrangian derivation of the above equation, due to the presence of the square root, some terms contain $\sqrt{R^2-6G}$ in the denominator, but this quantity is vanishing on the dS solution. An alternative and equivalent way to arrive at the same result is to verify that the condition \eqref{dsfg} is not satisfied.

For $\rho$ generic and for $k$ not vanishing, equation (\ref{bu}) appears intractable. Thus, at first order in the small parameter $\frac{1}{\rho_c}=\frac{\alpha^2}{12}$, we find
\be
3J^2=\left(\rho-\frac{ \rho ^2}{\rho_c}\right) \left(1- \frac{4 \,k }{\rho_c a^2}\right) \,. \label{bu1}
\ee
Let us take the simplest case $\omega=-1$, namely $\rho=\rho_0$. The solutions of the equation above read,
\begin{equation}
a(t)=\cosh[H_0 t]\,,\quad k=\mu^2 \,,\nonumber   
\end{equation}
\begin{equation}
a(t)=\sinh[H_0 t]\,,\quad k=-\mu^2 \,,   
\end{equation}
where
\begin{equation}
H_0^2=\frac{1}{3}\left(\rho_0-\frac{ \rho_0 ^2}{\rho_c}\right) \,,
\quad \mu^2=\frac{\left(\rho_0-\frac{ \rho_0 ^2}{\rho_c}\right)}{3}
\left[\left(\rho_0-\frac{ \rho_0 ^2}{\rho_c}\right)+1\right]^{-1}\,,
\end{equation}
where we have taken into account that $\left(\rho_0-\frac{ \rho_0 ^2}{\rho_c}\right)>0$.
Thus, for positive curvature $k$ we obtain a bounce solution, while for negative curvature we get a singular Big Bang solution at $t=0$. As in the cases of (\ref{sol00}) for mimetic gravity $J^2$ is not a constant and we do not have the dS solution.

For the general case $\omega\neq -1$ we observe that Eq. (\ref{bu1}) may be written as
\be
\dot{a}^2=A[a(t)]\,, 
\ee
with
\be
A(a)=\frac{1}{3} \left[ \rho_0 a^{-1-3\omega}-3k-\frac{ \rho_0^2}{\rho_c} a^{-4-6\omega}+\frac{4k\rho_0}{\rho_c}
\left(  \frac{\rho_0}{\rho_c}a^{-6(\omega+1)}-a^{1-3\omega  } \right) \right]\,.
\ee
Making an expansion around the critical point $A(a_*)=0$, we find, for small $t$, $a(t)\simeq a_*+A'_*\, t^2$.
As a result, when $a_*>0$, and $A'_* >0$, we may have finite regular FRWL cosmological solutions, as shown in the previous Section.

\subsection{Other Non-Polynomial gravity models}

Consider another example for any $k$, which generalizes the model firstly considered in Refs. \cite{aimeric,aimeric2} for $k=0$. The model is defined by the action
\be
I=\int d^4x \sqrt{-g}\left[\frac{R}{2}+\frac{\alpha}{6} \sqrt{(-\nabla_\mu R)^2}   \right]+I_m\,.
\ee
On a non flat FLRW, up to an integration by part, the gravitational Lagrangian reads
\be
L=-6a\dot{a}^2+6k a-6\alpha k \dot{a}-3\alpha  \dot{a}^3\,.
\ee
The variation with respect to $a$ with the matter contributions leads to the generalized second Friedmann equation,
\be
2 \dot{H}+3H^2+\frac{k}{a^2}+3\alpha H(H^2+\dot{H})=-p\,.
\ee
In the case $k=0$, and provided that $p=p_0$ constant, the dS solution in the flat patch exists. However, for
$k\neq 0$, the dS solutions appear problematic. This fact is confirmed by the first Friedmann equation in the case $k \neq 0$. As usual, by covariance, the Friedamnn equation can be obtained making use of matter conservation equation, and reads
\be
3J^2-3\alpha H^3=\rho\,,\quad J^2=H^2+\frac{k}{a^2}\,.
\ee
In the case $k=0$ with constant energy density $\rho=\rho_0$, the dS solution in the flat patch exists, but for $k \neq  0$ non-flat dS solutions do not exist.\\
\\
Another similar example contains the $G$ invariant, namely
\be
I=\int d^4x \sqrt{-g}\left[\frac{R}{2}+\frac{\alpha}{24} \sqrt{(-\nabla_\mu G)^2}   \right]+I_m\,.
\ee
The variation of the associated Lagrangian with respect to $a$ leads to the generalized second Friedmann equation
\be
2 \dot{H}+3H^2+\frac{k}{a^2}+3\alpha H^5+ \alpha \dot{ H}\left(5H^3+\frac{2 k}{a^3}\right)=-p\,,
\ee
from which it is possible to infer the first equation by making use of the matter conservation law,
\begin{equation}
3J^2+\alpha H \left(3H^4+\frac{2k}{a^3}\right)=\rho\,,\quad J^2=H^2+\frac{k}{a^2}\,.    
\end{equation}

Again, for $k=0$, the dS solution in the flat patch exists. But, for
$k \neq  0$, again,  the dS solution does not exist.

In these two cases, as for the previous $F(R,G)$ example involving a non analytical choice of $F$, we may understand the trouble with dS solution observing again that the non analicity present in the Lagrangians, in a covariant derivations, leads to variations containing denominators which are ill defined for the dS case. 

\section{Horndeski models}

We continue our investigation concerning Lagrangian models in FLRW space-times with another well known class of of scalar-tensor model admitting second order equations of motion in arbitrary space-times: the so called Hordenski models.

\subsection{An example of Horndeski model}

In this section we will investigate only a reduced sector of Horndeski in vacuum. The action of this sub-sector is
\be
I=\int d^4x \sqrt{-g}\left( \frac{R}{2}+\alpha G^{\mu \nu}\partial_\mu \phi\partial_\nu \phi -
  \frac{1}{2}\partial_\mu \phi \partial^{\mu} \phi- V(\phi) \right)\,.
\ee
This model has been studied without the potential in Refs. \cite{max,max2}. The associated Lagrangian, rewritten in coordinates using the metric \eqref{metricn}, reads
\be
L=-6 \frac{a\dot{a}^2}{N}+6kN a+6\alpha \dot{\phi}^2\left(\frac{ k a}{N}+ \frac{a \dot{a}^2}{N^3}\right)+
Na^3 \dot{\phi}^2-2Na^3 V(\phi)\,.
\ee
Making the variation with respect to $N(t)$, and considering $N=1$ after the variation, gives the generalized Friedmann equation
\be
3J^2-3\alpha  \dot{\phi}^2\left(J^2+2H^2 \right)=\frac{1}{2} \dot{\phi}^2+V(\phi)\,.
\label{f}
\ee
Moreover, performing the variation of the Lagrangian with respect to the field $\phi$ we obtain the equation of motion associated with $\phi$,
\be
\frac{d}{dt}\left[a^3(1+6\alpha J^2)\dot{\phi}  \right]=-a^3 \frac{d V}{d\phi}\,.\label{phi}
\ee
The above equations are in agreement with the non flat FLRW Hordenski equations in Ref. \cite{jap19}.

Note that performing an additional variation with respect to $a$ we can obtain the second generalized Friedmann equation. However, this equation simply follows from the two above equations \eqref{f} and \eqref{phi}.

When the potential is constant, i.e. $V=V_0$, we have
\be
\dot{\phi}=\frac{C}{a^3\left[1+6\alpha J^2\right]}\,,
\label{p}
\ee
where $C$ is a constant of integration.

We want to investigate the existence of dS solution. Firstly, we consider the $k=0$ case, and make the Ansatz  $a(t)=e^{H_0 t}$, $H=H_0=J_0$ and $J^2=J^2_0 = H_0^2$. 
Thus the equation of motion of $\phi$ becomes
\be
\dot{\phi}=\frac{C}{1+6\alpha H_0^2}e^{-3H_0 t}\,.
\ee
If $C=0$, and choosing $3H^2_0=V_0$, we have a flat dS-like solution.

On the other hand, if $C\neq 0$, we may satisfy the equation \eqref{f} imposing $\alpha <0$ and fixing
\be
H^2_0=-\frac{1}{18 \alpha}\, \quad V_0=\frac{1}{6 \alpha}\,.
\ee
For $k \neq 0$, the invariant, $J^2$   is not the equal to $H_0^2$ , and we have
\be
3J^2-3\alpha \dot{\phi}^2\left[J^2+2H^2 \right]=\frac{1}{2} \dot{\phi}^2+V_0\,.
\label{f12}
\ee
As a result, we still have the dS like solution when $C=0$. But for $C\neq 0$, the dS solution does not exist.
\\
\\

\subsection{Another example of Horndeski model}

Another example we propose is the following Horndeski model,
\begin{align}
I=\int_\mathcal M dx^4\sqrt{-g}\,&\biggl\lbrace\frac{R}{2}-\frac{\partial_\mu\phi\partial^\mu\phi}{2}+\alpha \sqrt{2|X|}R\nonumber\\
&\phantom{xx}+ \frac{\alpha}{\sqrt{2|X|}}\frac{|X|}{X}
[(\Box\phi)^2-(\nabla_\mu \nabla_\nu \phi)(\nabla^\mu \nabla^\nu \phi)]
\biggr\rbrace\,,\label{action}
\end{align}
which corresponds to
$G_4(\phi, X)=\alpha \sqrt{2|X|}$ with $2 X=-\partial_\mu\phi\partial^\mu\phi$ in the general formulation of Horndeski action. In what follows, we assume to deal with a real field, i.e. $\dot\phi^2>0$, in the FRW space-time. This model has been studied in Refs.~\cite{SebSSS,SebSSS2} in the SSS space-time and, when the potential of the field vanishes, admits the Reissner-Nordstrom solution, such that the $\alpha$ parameter plays the role of the charge. 
By assuming \eqref{metricn} the associated Lagrangian simply reads
\begin{equation}
L=-3 \frac{a\dot{a}^2}{N}+3kN +\frac{1}{2N}a^3 \dot{\phi}^2-Na^3 V(\phi)+6\alpha k a\dot\phi\,.
\end{equation}
The field equations are derived as
\begin{equation}
3 J^2=\frac{\dot\phi^2}{2}+V(\phi)\,,
\end{equation}
\begin{equation}
2(H^2+\dot H)+J^2+\frac{2\alpha k}{a^2}\dot\phi=-\frac{\dot\phi^2}{2}+V(\phi)\,,
\end{equation}
while the continuity equation for the field leads to
\begin{equation}
\ddot\phi+3H\dot\phi+V_{\phi}=-\frac{6\alpha H k}{a^2}\,,
\end{equation}
namely
\begin{equation}
\frac{d}{dt} (\dot\phi a^3+3\alpha k a^2)=-\frac{d V}{d\phi} a^3\,.
\end{equation}
Clearly, in the case of $k=0$ we recover GR, but when $k\neq 0$ a new contribution appears in the field equations. When $V=0$ we have
\begin{equation}
\dot\phi=\frac{C_0}{a^3}-\frac{3\alpha k}{a}\,,
\end{equation}
and when $C_0=0$ we get the dS solution only if $k=0$.

\section{Lovelock gravity and its variants}

In this section, after a brief review of Lovelock gravity, we will show that the problematic issue associated with the dS solution in all dS patches is not present within this model.

\subsection{Review of Lovelock gravity}

The Lovelock gravity action in the case $D$ $(\geq 4)$ reads
\begin{align}
\label{action_lov}
I=&\frac{1}{16\pi G_N}\int dx^D\sqrt{-g}\sum_{p=0}^{[n/2]}\alpha_{(p)}{\mathcal{L}}_{(p)}+I_{\rm matter},\\
{\mathcal{L}}_{(p)}:=&\frac{1}{2^p}\delta^{\mu_1\cdots \mu_p\nu_1\cdots \nu_p}_{\rho_1\cdots \rho_p\sigma_1\cdots \sigma_p}R_{\mu_1\nu_1}^{\phantom{\mu_1}\phantom{\nu_1}\rho_1\sigma_1}\cdots R_{\mu_p\nu_p}^{\phantom{\mu_p}\phantom{\nu_p}\rho_p\sigma_p}\,.
\end{align}
The $\delta$ stands for the totally anti-symmetric products of the Kronecker deltas, normalized to take values $0$ and $\pm 1$~\cite{Lov}, and it is defined by
\begin{equation}
\delta^{\mu_1\cdots \mu_p}_{\rho_1\cdots \rho_p}:=p!\delta^{\mu_1}_{[\rho_1}\cdots \delta^{\mu_p}_{\rho_p]}.
\end{equation}
The $\alpha_{(p)}$ are coupling constants with dimension $({\rm length})^{2(p-1)}$. We can choose  $\alpha_{(0)}=-2\Lambda$, where $\Lambda$ is the cosmological constant.

Explicitly, the first terms of the Lovelock gravity action are
\begin{align}
{\mathcal{L}}_{(0)}:=& 1\,,\nonumber\\
{\mathcal{L}}_{(1)}:=& R\,,\nonumber\\
{\mathcal{L}}_{(2)}:=& R^2-4R_{\mu\nu}R^{\mu\nu}+R_{\mu\nu\rho\sigma}R^{\mu\nu\rho\sigma}\,.
\end{align}
In even dimensions, the contributions to the action of the $D/2$-th order and above Lagrangians is null, as these Lagrangians become topological invariant and does not contribute to the field equations. In other words, the variation of these terms with respect to the metric gives a total derivative which does not contribute to the equations of motion.

\noindent For instance, for $D=4$, ${\mathcal{L}}_{(2)}=G$ is the Gauss-Bonnet topological invariant which does not contribute to the field equations. Moreover, the higher order terms do not contribute to the field equations. Therefore the action reduces to Einstein-Hilbert action plus a cosmological constant.

The equation of motions for any $D$ $(\geq 4)$ are 
\begin{equation} 
\mathcal{G}_{\mu\nu}=8 \pi G_N {T}_{\mu\nu}\,, \label{beqL}
\end{equation} 
where ${T}^\mu_{~~\nu}$ is the energy-momentum tensor for matter fields obtained from $I_{\rm matter}$ and
\begin{align} 
\mathcal{G}_{\mu\nu} :=& \sum_{p=0}^{[n/2]}\alpha_{{(p)}}{G}^{(p)}_{\mu\nu}\,, \label{generalG}\\
{G}^{\mu(p)}_{~~\nu}:=& -\frac{1}{2^{p+1}}\delta^{\mu\eta_1\cdots \eta_p\zeta_1\cdots \zeta_p}_{\nu\rho_1\cdots \rho_p\sigma_1\cdots \sigma_p}R_{\eta_1\zeta_1}^{\phantom{\eta_1}\phantom{\zeta_1}\rho_1\sigma_1}\cdots R_{\eta_p\zeta_p}^{\phantom{\eta_p}\phantom{\zeta_p}\rho_p\sigma_p}\,.
\end{align} 
The tensor ${G}^{(p)}_{\mu\nu}$ is given from ${\mathcal{L}}_{(p)}$.
${G}^{(p)}_{\mu\nu}\equiv 0$ is satisfied for $p\ge [(D+1)/2]$.
Note that we obtain
\begin{align}
G^{\mu(p)}_{~~\mu} \equiv  \frac{2p-D}{2}\,{\cal L}_{(p)}.\label{id-L}
\end{align}
It is easy to show that the equations of motion are second order partial differential equations, since only curvature tensors appear. In fact, we have
\begin{align}
{G}^{(0)}_{\mu\nu} =& -\frac12 g_{\mu\nu}\,,\nonumber\\
{G}^{(1)}_{\mu\nu}=&\, R_{\mu\nu}-\frac12 Rg_{\mu\nu}\,,\nonumber\\
{G}^{(2)}_{\mu\nu}=&\, 2\biggl(RR_{\mu\nu}-2R_{\mu\rho}R^\rho_{\phantom{\rho}\nu}-2R^{\rho\sigma}R_{\mu\rho\nu\sigma}+R_{\mu}^{\phantom{\mu}\rho\sigma\gamma}R_{\nu\rho\sigma\gamma}\biggl)-\frac12 g_{\mu\nu}{\mathcal{L}}_{(2)}\,.
\end{align}

\subsection{Lovelock gravity in non flat FLRW space-times}

The first Lovelock-Friedmann (LF) equation of motion in a generic FLRW space-times reads \cite{Faria}
\be
8\pi G_N T_{00}= \sum_{p=0}^{[D/2]}\beta_{{(p)}}{J^2(t)}^{(p)},\,
\label{lj}
\ee
where the constant coefficients $\beta_{p}$ are proportional to the $\alpha_p$. The other equation can be derived by the above equation, as in GR, by  taking into account of stress energy matter conservation
\be
\dot{\rho}+3H (\rho+p)=0\,.
\ee
A remark is in order. The LF equation depends only on the SSS invariant $J^2(t)=H^2(t)+k/a^2(t)$. As a consequence, when there is a non-vanishing positive cosmological constant or a constant matter density, there exists the de Sitter (dS) solution, namely the solutions for the three dS patches corresponding to $k=0$, $k>0 $ and $k<0$ are explicit solutions of the LF equation.

Obviously, in $D=4$ the situation is trivial, since Lovelock gravity reduces to GR plus cosmological constant. We will come back to Lovelock gravity in the next subsection, but in a different contest.

\subsection{A degenerate Lovelock gravity in $D=4$}

As already mentioned, in $D=4$ the Lovelock gravity is essentially GR. In order to go beyond GR in $D=4$, we make use of an approach, discussed in Ref. \cite{zerbini13} and reference quoted therein. The key idea is to  start  considering a Gauss-Bonnet-Lovelock gravity initially in $D$ dimension.  Then, making use of a particular choice of the Gauss-Bonnet coupling, we may take the limit $D \rightarrow 4$. Recently the issue has been reconsidered in Ref. \cite{Gla}.

The starting point is the following particular modified $F(R,G)$-model in $D>4$ which belong to Lovelock class, 
\be
I=\frac{1}{16\pi G_N}\int d^Dx \sqrt{-g}\left( \frac{R}{2}-2\Lambda-\frac{\beta}{D-4}\,G\right) +I_m\,.
\ee
Thus, working  in a non flat D-dimensional FLRW space time and then taking the limit  $D \rightarrow 4$, we obtain
\be
3J^2(t)-\Lambda-3\beta J^4(t)= \rho\,,
\ee
which generalizes the equation obtained in Ref. \cite{zerbini13} for $k=0$. In this degenerate Gauss-Bonnet-Lovelock model, the dS solution exists in all patches, as soon as $\rho$ is constant or vanishing.

For  generic $\rho$, we find
\be
J^2=\frac{1}{2 \beta}\left[1-\sqrt{1-\frac{4\beta}{3}(\Lambda+ \rho)}  \right]\,.
\ee
Thus, there exist bounce solutions for this model in FLWR space-times. 

\section{Conclusions}

In this paper, we have investigated some extended or modified gravitational models in a generic non flat FLRW space-times. We have shown that, apart from GR, Lovelock gravity and reduced Lovelock gravity, the generalized Friedmann equations related to these modified extended models admits pseudo dS solutions, namely  only  for the flat FLRW case, the metrics look like the flat dS patch, but for positive and negative spatial curvature, the related solutions are not dS ones. In some cases, this issue  has been identified with a singularity in the variational procedure, due to the non analyticity in the Lagrangians.

One of the motivations to deal with such models is the fact that the generalized Friedmann equation for $k=0$ admits as a solution a cosmic bounce, a non singular cosmological solution, which avoids the Big Bang initial singularity present in  GR. We presented an effective Lagrangian derivation based essentially on a mimetic extended approach and a modified gravity approach based on a specific choice of $F(R, G)$ model.

The modified Friedmann equations obtained by these two approaches are quite different, they however give the same modified loop-cosmology Friedamnn equation in flat FLRW space-time. The presence of non vanishing FLRW spatial curvature drastically changes the modified Friedmann equation, and dS space-time is not a solution of these equations. In some cases, when $k <0$, and $\omega=-1$, a Big Bang singularity is present.  

Thus, in order to ``solve'' this pseudo dS issue, it is tempting to make the following Ansatz. Let us consider as  effective Friedmann equation the following one
\be
6J^2(t)-f(J^2)- J\frac{\partial f}{\partial J} =2  \rho\,.
\label{?}
\ee
As a consequence, this equation depends only on the invariant $J^2$. As usual, making the choice 
\begin{equation}
 f(J^2)=6J^2+\frac{12}{\alpha^2}\left[
 1-\sqrt{1-\alpha^2 J^2}-\alpha J\arcsin\left[\alpha J\right]
 \right]\,,
\end{equation}
we arrive at the modified Friedmann equation
\begin{equation}
3J^2=\rho\left(1-\frac{\rho}{\rho_c}\right)\,,\qquad \text{where}\qquad  \rho_c=\frac{12}{ \alpha^2}\,.
\label{J}
\end{equation}
This should be the generalization of the $k=0$ QLC modified Friedmann equation, valid for non vanishing curvature, since this equation, for $k=0$ reduces to the one already known, and the above equation, admits dS solutions for all patches. Furthermore, for $\alpha$ small, gives corrections to the GR, and may  admit regular bounce solutions.

This may be seen for specific choices for the barotropic parameter $\omega$. In fact, for $\omega=-1/3$, we obtain the symmetric bounce solution
\be
y=a(t)^2=\left[ \frac{\rho_0^2}{\rho_c(\rho_0-3k)}+\frac{(\rho_0-3k) t^2}{3} \right]\,,
\ee
where we assume $\rho_0 > 3k$. 

Another choice is $\omega=-\frac{2}{3}$. In this case, the symmetric bounce reads 
\be
y(t)=a(t)^2=\left( \frac{\rho_0}{\rho_c}+\frac{3k}{\rho_0}+
\frac{\rho_0\, t^2}{12} \right)^2\,.
\ee
in which $\frac{\rho_0}{\rho_c}+\frac{3k}{\rho_0}>0$. Note that in this case, we have a symmetric bounce also in GR as soon as $k>0$. 

For $0< \omega <1 $,
it is not easy to have an exact and explicit solution, but we may again start separating the variable in equation (\ref{J}) with $y=a^2$
\be
\int \frac{dy}{\sqrt{Y(y)}}=t\,,
\ee
where we have used the matter conservation law and put
\be
Y(y)=\frac{4 \rho_0}{3}\left[y^{1/2-3\omega/2}-\frac{\rho_0}{\rho_c}\,y^{-(1+3\omega)}-\frac{3k\,y}{\rho_0} \right]\,.
\ee
If $k$ is not vanishing, this integral is only exactly computable only for $\omega=-1/3$ and $\omega=-2/3$, as reported above. 

Furthermore, making an expansion around the critical point $ Y(y_*)=0$, namely
\be
y_*^{3/2+3\omega/2}-\frac{3k}{\rho_0}y_*^{2+3\omega}=
\frac{\rho_0}{\rho_c}\,,
\label{J3}
\ee
we obtain again the approximate solution valid  for small $t$
\be
y(t)\simeq y_*+\left[\frac{(1+\omega)\rho_0}{2}\,y_*^{-1/2-3\omega/2}-k(2+3\omega)   \right]\,t^2\,.
\label{J4}
\ee
Above, $y_*$ is solution of the transcendental equation 
(\ref{J3}).  It is easy to show that for $\omega=-2/3\,, -1/3$, we get the approximate solution related to the exact solution found before. Furthermore, for $k=0$, we obtain an approximate result compatible with the exact well known solution. 

Thus, from equation (\ref{J4}), we conclude that we have to deal with a regular symmetric bounce as soon as $y_*>0$ and  
\be
\left[\frac{(1+\omega)\rho_0}{2}\,y_*^{-1/2-3\omega/2}-k(2+3\omega)   \right]\,>0\,.
\label{J5}
\ee
For example, for $\rho \ll \rho_c$ and for $k>0$, we obtain a regular symmetric bounce.

Finally, the solutions we found can be roughly constrained with data coming from the Planck experiment \cite{Planck2}. In particular, we can estimate the value of the critical density $\rho_c$, in the inflationary cases only. In fact, we know, from the Planck experiment, the upper bound of the Hubble parameter during inflation (in our units system)
\begin{equation}
    H_*^2 \lesssim 10^{-10}\,.
\end{equation}
Therefore, since $\rho_c \propto 1 / \alpha^2$, and in our models we always require, from the proposed definition of $f(H)$, $H^2 \lesssim 1/\alpha^2$, we can make a rough estimation of $\rho_c \gtrsim 10^{-10}$. To obtain more precise constraints, in principle we should check our results against the main inflation perturbation observables (spectral indices, tensor to scalar ratio, etc..). However, this analysis goes beyond the scope of this paper and might be an interesting starting point for future works.


\begin{thebibliography}{99}

\bibitem{Plank1}
  R.~Adam {\it et al.} [Planck Collaboration],
  Astron.\ Astrophys.\  {\bf 594} (2016) A1.
 
 \bibitem{Planck2}
 Y.~Akrami \textit{et al.} [Planck],
``Planck 2018 results. X. Constraints on inflation,''
[arXiv:1807.06211 [astro-ph.CO]].
 
\bibitem{Plank2}
  P.~A.~R.~Ade {\it et al.} [Planck Collaboration],
  Astron.\ Astrophys.\  {\bf 594} (2016) A13.

\bibitem{DiVa}
  E.~Di Valentino, A.~Melchiorri and J.~Silk,
  Nat.\ Astron.\  (2019).

\bibitem{Lov} 
  D.~Lovelock,
  J.\ Math.\ Phys.\  {\bf 12}, 498 (1971).

\bibitem{Bo} 
  M.~Bojowald,
  Phys.\ Rev.\ Lett.\  {\bf 86}, 5227 (2001).
  
\bibitem{Bo2}
  M.~Bojowald,
  Living Rev.\ Rel.\  {\bf 11}, 4 (2008).
  
\bibitem{Bo19} 
  M.~Bojowald,
  arXiv:1906.03146 [gr-qc].

\bibitem{da19}
  M.~Assanioussi, A.~Dapor, K.~Liegener and T.~Paw{\l}owski,
  Phys.\ Rev.\ D {\bf 100} (2019) no.8,  084003.


\bibitem{noi} 
  R.~Di Criscienzo, S.~A.~Hayward, M.~Nadalini, L.~Vanzo and S.~Zerbini,
  Class.\ Quant.\ Grav.\  {\bf 27}, 015006 (2010).
  
 \bibitem{noi2} 
  G.~Acquaviva, R.~Di Criscienzo, M.~Tolotti, L.~Vanzo and S.~Zerbini,
  Int.\ J.\ Theor.\ Phys.\  {\bf 51}, 1555 (2012).
  
\bibitem{noi1} 
  S.~A.~Hayward, R.~Di Criscienzo, L.~Vanzo, M.~Nadalini and S.~Zerbini,
  Class.\ Quant.\ Grav.\  {\bf 26}, 062001 (2009).
  
\bibitem{zocc} 
  R.~Di Criscienzo, M.~Nadalini, L.~Vanzo, S.~Zerbini and G.~Zoccatelli,
  Phys.\ Lett.\ B {\bf 657}, 107 (2007).

\bibitem{horn} 
  G.~W.~Horndeski,
  Int.\ J.\ Theor.\ Phys.\  {\bf 10}, 363 (1974).
  
\bibitem{muk} 
  A.~H.~Chamseddine and V.~Mukhanov,
  JHEP {\bf 1311}, 135 (2013).
  
\bibitem{muk1} 
  A.~H.~Chamseddine, V.~Mukhanov and A.~Vikman,
  JCAP {\bf 1406}, 017 (2014).

\bibitem{vik} 
  L.~Mirzagholi and A.~Vikman,
  JCAP {\bf 1506}, no. 06, 028 (2015).

\bibitem{Derue} 
  N.~Deruelle and J.~Rua,
  JCAP {\bf 1409}, 002 (2014).

\bibitem{barvi} 
  A.~O.~Barvinsky,
  JCAP {\bf 1401}, no. 01, 014 (2014).

\bibitem{beke} 
  J.~D.~Bekenstein,
  Phys.\ Rev.\ D {\bf 48}, 3641 (1993).

\bibitem{mata} 
 F.~Arroja, N.~Bartolo, P.~Karmakar and S.~Matarrese,
  JCAP {\bf 1509}, 051 (2015).
 
\bibitem{golo} 
  A.~Golovnev,
  Phys.\ Lett.\ B {\bf 728}, 39 (2014).



\bibitem{Lim} 
  E.~A.~Lim, I.~Sawicki and A.~Vikman,
  JCAP {\bf 1005}, 012 (2010).

\bibitem{Capo} 
  S.~Capozziello, J.~Matsumoto, S.~Nojiri and S.~D.~Odintsov,
  Phys.\ Lett.\ B {\bf 693}, 198 (2010).


\bibitem{Odi1} 
  S.~Nojiri and S.~D.~Odintsov,
  Mod.\ Phys.\ Lett.\ A {\bf 29}, no. 40, 1450211 (2014).

\bibitem{Odi2} 
  J.~Matsumoto, S.~D.~Odintsov and S.~V.~Sushkov,
  Phys.\ Rev.\ D {\bf 91}, no. 6, 064062 (2015).
\bibitem{Odi3} 
  S.~D.~Odintsov and V.~K.~Oikonomou,
  arXiv:1508.07488 [gr-qc].


\bibitem{myrza1}
  R.~Myrzakulov, L.~Sebastiani and S.~Vagnozzi,
  Eur.\ Phys.\ J.\ C {\bf 75} (2015) 444.
  
\bibitem{myrza2}
  M.~Raza, K.~Myrzakulov, D.~Momeni and R.~Myrzakulov,
  Int.\ J.\ Theor.\ Phys.\  {\bf 55} (2016) no.5,  2558.
  
\bibitem{myrza3}
  R.~Myrzakulov, L.~Sebastiani, S.~Vagnozzi and S.~Zerbini,
  Fund.\ J.\ Mod.\ Phys.\  {\bf 8} (2015) 119.
  
\bibitem{mim_sv1}
  A.~Casalino, M.~Rinaldi, L.~Sebastiani and S.~Vagnozzi,
  Phys.\ Dark Univ.\  {\bf 22} (2018) 108.
  
\bibitem{mim_sv2}
  A.~Casalino, M.~Rinaldi, L.~Sebastiani and S.~Vagnozzi,
  Class.\ Quant.\ Grav.\  {\bf 36} (2019) no.1,  017001.



\bibitem{Deffa} 
  C.~Deffayet, X.~Gao, D.~A.~Steer and G.~Zahariade,
  Phys.\ Rev.\ D {\bf 84}, 064039 (2011).

\bibitem{DeFe} 
  A.~De Felice, T.~Kobayashi and S.~Tsujikawa,
  Phys.\ Lett.\ B {\bf 706}, 123 (2011).


\bibitem{aimeric}
  S.~Chinaglia, A.~Colleaux and S.~Zerbini,
  Galaxies {\bf 5} (2017) no.3,  51.
  
\bibitem{aimeric2}
  A.~Colleaux, S.~Chinaglia and S.~Zerbini,
  Int.\ J.\ Mod.\ Phys.\ D {\bf 27} (2018) no.03,  1830002.
  
\bibitem{NO} 
  S.~Nojiri and S.~D.~Odintsov,
  Phys.\ Lett.\ B {\bf 691}, 60 (2010).

\bibitem{Cogno11} 
  G.~Cognola, E.~Elizalde, L.~Sebastiani and S.~Zerbini,
  Phys.\ Rev.\ D {\bf 83}, 063003 (2011).

\bibitem{Muk3} 
  A.~H.~Chamseddine and V.~Mukhanov,
  JCAP {\bf 1703}, no. 03, 009 (2017).

\bibitem{Muk4} 
  A.~H.~Chamseddine, V.~Mukhanov and T.~B.~Russ,
  arXiv:1912.03162 [hep-th].


\bibitem{Lang} 
  D.~Langlois, H.~Liu, K.~Noui and E.~Wilson-Ewing,
  Class.\ Quant.\ Grav.\  {\bf 34}, no. 22, 225004 (2017).

\bibitem{Haro} 
  J.~de Haro, L.~Arest\'e Sal\'o and S.~Pan,
  Gen.\ Rel.\ Grav.\  {\bf 51}, no. 4, 49 (2019).
  
  
\bibitem{Biswas1} 
  T.~Biswas, T.~Koivisto and A.~Mazumdar,
  JCAP {\bf 1011}, 008 (2010).
\bibitem{Biswas2} 
  T.~Biswas, A.~S.~Koshelev, A.~Mazumdar and S.~Y.~Vernov,
  JCAP {\bf 1208}, 024 (2012).


\bibitem{superSeba}
R.~Myrzakulov, L.~Sebastiani and S.~Zerbini,
  Gen.\ Rel.\ Grav.\  {\bf 45}, 675 (2013).
  
\bibitem{Monica} 
  G.~Cognola, M.~Gastaldi and S.~Zerbini,
  Int.\ J.\ Theor.\ Phys.\  {\bf 47}, 898 (2008),

\bibitem{Gao} 
  C.~Gao,
  Phys.\ Rev.\ D {\bf 86}, 103512 (2012).



\bibitem{Myrza4} 
  R.~Myrzakulov and L.~Sebastiani,
  Astrophys.\ Space Sci.\  {\bf 352}, 281 (2014).


  \bibitem{jap19} 
  S.~Akama and T.~Kobayashi,
  Phys.\ Rev.\ D {\bf 99}, no. 4, 043522 (2019).



   
\bibitem{Faria} 
  N.~Dereulle and L. Faria-Busto,  
  Phys.\ Rev.\ D {\bf 41}, 3696 (1990).

  
\bibitem{zerbini13} 
  G.~Cognola, R.~Myrzakulov, L.~Sebastiani and S.~Zerbini,
  Phys.\ Rev.\ D {\bf 88}, no. 2, 024006 (2013).


\bibitem{Gla} 
  D.~Glavan and C.~Lin,
  arXiv:1905.03601 [gr-qc].
  
\bibitem{max}
  M.~Rinaldi,
  Phys.\ Dark Univ.\  {\bf 16} (2017) 14.
  
\bibitem{max2}
  A.~Casalino and M.~Rinaldi,
  Phys.\ Dark Univ.\  {\bf 23} (2019) 100243.
  
  \bibitem{SebSSS}
E.~Babichev, C.~Charmousis and A.~Leh\'ebel,
  JCAP {\bf 1704}, 027 (2017).
  
   \bibitem{SebSSS2}
L.~Sebastiani,
  Int.\ J.\ Geom.\ Meth.\ Mod.\ Phys.\  {\bf 15}, no. 09, 1850152 (2018).
  
  
\end{thebibliography}
\end{document}